\vsize=23.5truecm \hsize=16truecm
\baselineskip=0.5truecm \parindent=0truecm
\parskip=0.2cm \hfuzz=1truecm

\font\scap=cmcsc10

\font\tenmsb=msbm10
\font\sevenmsb=msbm7
\font\fivemsb=msbm5
\newfam\msbfam
\textfont\msbfam=\tenmsb
\scriptfont\msbfam=\sevenmsb
\scriptscriptfont\msbfam=\fivemsb

\newcount\eqnumber
\eqnumber=1
\def\neweq{{\rm{(\the\eqnumber)}}\global\advance\eqnumber by 1}
\def\eqdef#1{\eqno\xdef#1{\the\eqnumber}\neweq}
\def\newaeq{{\rm{(\the\eqnumber a)}}\global\advance\eqnumber by 1}
\def\eqdaf#1{\eqno\xdef#1{\the\eqnumber}\newaeq}
\def\eqdisp#1{\xdef#1{\the\eqnumber}\neweq}
\def\eqdasp#1{\xdef#1{\the\eqnumber}\newaeq}

\newcount\refnumber
\refnumber=1
\def\newref{{\the\refnumber}\global\advance\refnumber by 1}
\def\refdef#1{{\xdef#1{\the\refnumber}}\newref}

\newcount\fignumber
\fignumber=1
\def\newfig{{\the\fignumber}\global\advance\fignumber by 1}
\def\figdef#1{{\xdef#1{\the\fignumber}}\newfig}

\def\smallskip{\vskip 3pt}
\def\medskip{\vskip 6pt}
\def\bigskip{\vskip 12pt}

\def\bol{\scalebox{0.7}{$\bullet$}}

\input graphicx

\centerline{\bf Deautonomising the Lyness mapping}
\bigskip
\medskip{\scap B. Grammaticos} and {\scap A. Ramani}
\quad{\sl Universit\'e Paris-Saclay and Universit\'e de Paris-Cit\'e, CNRS/IN2P3, IJCLab, 91405 Orsay, France}
\medskip{\scap R. Willox} \quad
{\sl Graduate School of Mathematical Sciences, the University of Tokyo, 3-8-1 Komaba, Meguro-ku, 153-8914 Tokyo, Japan }
\bigskip
{\sl Abstract}
\smallskip
We examine the Lyness mapping (an integrable $N$th-order discrete system which can be generated from a one-dimensional reduction of the Hirota-Miwa equation) from the point of view of deautonomisation. We show that only the $N=2$ case can be deautonomised when one works with the standard form of the mapping. However it turns out that deautonomisation is possible for arbitrary $N$ when one considers the derivative form of the Lyness mapping. The deautonomisation of the derivative of the $N=2$ case leads to a result we have never met before: the secular dependence in the coefficients of the mapping enters through two different exponential terms instead of just a single one. As a consequence, it turns out that a limit of this multiplicative dependence towards an additive one is possible without modifying the dependent variable. Finally, the analysis of the `late' singularity confinement of the $N=2$ case leads to a novel realisation of the full-deautonomisation principle: the dynamical degree is not given (as is customary) simply by the solution of some linear or multiplicative equation, but is present in the growth of the non-linear (and non-integrable) late-confinement conditions.
\bigskip
1. {\scap Introduction}
\medskip
While the beginning of the modern era of integrability in differential systems can unarguably be associated with the discovery of the soliton by Kruskal and Zabusky [\refdef\kruskal], this is less clearcut when it comes to difference systems. Still, it is indisputable that the derivation of what came to be known as the Quispel-Roberts-Thompson (QRT) mapping [\refdef\qrt] did play an important role in launching the activity in the domain of discrete systems that we have seen over the last 30 years. To be sure, the fundamentals of the approach, as Veselov points out in [\refdef\veselov], can be traced back to Euler with the parametrisation of the bi-quadratic relation
$$\alpha x^2y^2+\beta xy(x+y)+\gamma (x^2+y^2)+\epsilon xy+\zeta(x+y)+\mu=0,\eqdef\zena$$
in terms of elliptic functions. (The latter is actually given in Baxter's book [\refdef\baxter] without any attribution or references, which suggests that by then the result was already long part of the common heritage). Still, the derivation of the QRT mapping, of which (\zena) is the invariant, was the crucial step establishing the link with integrability studies. Moreover, soon after proposing a mapping that has (\zena) as invariant, for general values of the parameters, the authors of [\qrt] presented an extension to what came to be called the {\sl asymmetric} case, associated to the invariant
$$\alpha x^2y^2+\beta x^2y+\delta xy^2+\gamma x^2+\kappa y^2+\epsilon xy+\zeta x+\lambda y+\mu=0.\eqdef\zdyo$$
(It is interesting that while the integration of (\zena) was known for over a century, that of (\zdyo) had to wait for more than a decade. In the end it turned out  that it can be parametrised in terms of elliptic functions [\refdef\autoqrt,\refdef\iatrou], just as (\zena)).

Be that as it may, the existence of whole families of integrable mappings provided a rich playground for the proposal and assay of various techniques, foremost of which were those designed to detect the integrable character of a given mapping of the plane. While studying the singularities that may appear spontaneously in such systems, it was observed that when the system was integrable by spectral methods, any singularity appearing due to some special choice of initial conditions did disappear after a certain number of iteration steps. This property was dubbed `singularity confinement' [\refdef\sincon] and it was readily elevated to a discrete integrability criterion. Here two important caveats are necessary. First, there exists a whole class of discrete systems, integrable through linearisation [\refdef\tremblay], which can possess unconfined singularities but this does not compromise their integrable character. Second, a whole class of non-integrable mappings with only confined singularities also exists [’\refdef\hiv,\refdef\diller]. Does this signal the inadequacy of singularity confinement as an integrability criterion? Nothing is further from the truth [\refdef\redemp], as we shall see shortly, but for this we first need to introduce the notion of `deautonomisation'.

A domain where singularity confinement excels is that of deautonomisation [\refdef\royal]. The latter consists in deriving non-autonomous extensions of initially autonomous, integrable, mappings, by assuming that the various parameters appearing in them can be functions of the independent variable. Using the singularity confinement criterion in order to obtain the precise forms of said functions then necessarily results into an integrable non-autonomous system [\royal]. In fact it was the deautonomisation approach that provided the key to resolving the paradox of non-integrable mappings with confined singularities. It was first noticed [\refdef\redeem], and subsequently rigorously proven [\refdef\stokes], that the deautonomisation based on singularity confinement provides a definite way to rule on the integrability of a given system, leading to what was termed the `full-deautonomisation' method as a discrete integrability criterion. 

The mechanism that underlies the confinement of singularities in the case of birational mappings (which constitute the vast majority of those studied) is based on simplifications of the rational expressions  for the iterates of the mapping, leading to the disappearance of any indefiniteness in those iterates. These simplifications also have a consequence that gave birth to another discrete integrability criterion. Studying the degrees of the polynomials (in the variables introduced by the initial conditions) in the numerator and denominator of the iterates of the mapping, it was observed that their degree increased slowly (in fact, polynomially) in the case of integrable mappings, while for non-integrable ones the growth was exponential [\diller]. This growth can be assessed quantitatively by considering what is called the dynamical degree. The latter is obtained from the degree $d_n$ of the iterates, for some generic initial conditions, as
$$\lambda=\lim_{n\to\infty}d_n^{1/n}.\eqdef\ztri$$
It is greater than 1 when the growth is exponential and equal to 1 otherwise [\diller].
The full-deautonomisation approach mentioned above originated in the observation that the conditions for deautonomisation of some autonomous mapping, be it integrable or not, yield direct information on the dynamical degree. When the non-autonomous mapping obtained from the deautonomisation is integrable this amounts to the trivial observation that the dynamical degree inferred from the deautonomisation conditions is equal to 1. It is in the non-integrable case however that things become truly interesting, as the confinement conditions lead to the exact value of the dynamical degree for both the non-autonomous map as well as for the original autonomous one.

The main bulk of integrability studies for one-dimensional discrete systems concerns second-order mappings. Over the past decades an enormous effort has gone into developing methods for the study of these low-order systems. And thus, naturally, the question arose whether these methods can still be of help when venturing into what is largely a {\sl terra incognita} of higher-order systems. In [\refdef\higher] we offered some insight into how one can adapt the existing methods in order to deal with systems of orders 3 or higher. We showed that the study of `singularities' and of `growth' is again an essential tool for the assessment of the integrable character of a given higher order mapping. However, both notions, unambiguously defined in the case of second order, need some clarification in the higher-order case. In all cases of order $N$ studied, it turned out that we could take generic initial conditions for the dependent variables $x_0, x_1, \cdots,x_{N-2}$ and compute the degree of the successive iterates in terms of $x_{N-1}$. Similarly, in our singularity analysis, it was sufficient to restrict ourselves to a situation where $x_{N-1}$ was the only initial condition allowed to take special values leading to singularities. It is our empirical finding that this approach works and leads to the proper answer; Still, it is clear that it is devoid of a rigorous foundation, the results obtained in [\diller] being specific to the second-order case. With this point clarified we can now proceed to the study of the Lyness mapping.
\bigskip
2. {\scap The Lyness mapping}
\medskip
This paper focuses on the Lyness mapping [\refdef\lyness,\refdef\lynesstoo,\refdef\kocicetal] and its non-autonomous extensions. The Lyness system is particularly interesting because it is a $N$th-order mapping, integrable at all orders in a sense which will be made more precise below (see also [\refdef\tranetal] for a construction of sufficiently many first integrals and a conjecture that the Lyness map is in fact Liouville integrable at all orders). The autonomous form of the map is usually written as:
$$ x_{n+N}x_n=a+  x_{n+1} +x_{n+2}+\dots+ x_{n+N-1},\eqdef\ztes$$
where $N\geq2$. As reported by Lyness in [\lyness] and [\lynesstoo], when $a=1$, this mapping is in fact periodic with period $3N-1$ when $N= 2$ or 3. Although this property seems to break down for all $N\geq4$ (see e.g. [\kocicetal] and [\refdef\groveladas]), we shall nevertheless impose $a\neq1$ in (\ztes) for all $N\geq2$.

As we shall see it is also interesting to study (\ztes) in its derivative form, which is the following $N+1$st order mapping:
$$ x_{n+N}(1+x_n)=x_{n+1}(1 +x_{n+N+1}).\eqdef\zpen$$
(When integrating (\zpen) once, so as to obtain (\ztes), the degree of freedom in the initial conditions for (\zpen) that one loses compared to (\ztes), is compensated by the parameter $a$ which plays the role of an integration constant.)

In order to assess the integrability of the Lyness mapping we start by considering its singularity structure.
The simplest case is that for $N=2$,
$$ x_{n+2}x_n=a+ x_{n+1}, \eqdef\zhex$$
which is a well-known integrable mapping belonging to the QRT family [\qrt]. 
We shall not go into a detailed singularity analysis of this mapping and therefore omit the (cyclic) singularity pattern that arises from $x_{n+1}=\infty$ (the reader is referred to [\refdef\lynett] or [\higher] for more detail). The only singularity that is relevant to our further discussion is that arising from $x_{n+1}=-a$ (for generic $x_n$) for which we obtain  the confined pattern $\{ -a, 0,-1,\infty,\infty,-1,0, -a\}$. 
Similarly, for the case $N=3$,
$$ x_{n+3}x_n=a+  x_{n+1}+ x_{n+2}, \eqdef\zhep$$
for generic $x_n$ and $ x_{n+1}$, the sole singularity that is of interest to us arises at $ x_{n+2}=-a- x_{n+1}\equiv x_*$. This singularity again confines, but one only recovers full (2 dimensional) freedom in terms of $x_n$ and $x_{n+1}$ at the iterate $x_{n+13}$, which yields the confined singularity pattern $\{ x_*, 0,-1,\bol,\infty,\infty,\bol,-1,0,*,*\}$, where  $\bol$ is used to indicate finite expressions involving $a$ and the free initial conditions  (here $x_n$ and $x_{n+1}$), and $*$ to indicate a succession of such functions that are not functionally independent as functions of the (free) initial conditions. 

A similar singularity analysis can be easily carried out for several higher values of $N$. In every case we obtain for the singularity at $x_{n+N-1}=x_*\equiv a-(x_n+\cdots+x_{n+N-2})$ a confined singularity pattern of the form:
 $\{x_*, 0,-1,(N-2 {\rm\ finite\ values\ \bol}),\infty,\infty,(N-2 {\rm\ finite\ values\ \bol}),-1,0, (N-1 {\rm\ finite\ values}\ *)\}$. We consider it safe to surmise that this holds for  all values of $N$, i.e. all higher-order Lyness mappings possess confined singularities of this type. As explained in detail in [\higher], assuming this to be the case, the express method for assessing degree growths for second order mappings that we introduced in [\refdef\express] then suggests that the dynamical degree of the Lyness mapping (\ztes), for general $N$, should be given by the greatest real root of the characteristic polynomial
 $$1-\lambda - \lambda^{2N}+\lambda^{2N+1} = (\lambda-1)(\lambda^{2N}-1),\eqdef\charpolyla$$
 that can be deduced from the above singularity pattern. This root being equal to 1 suggests that the mapping should be integrable for all $N\geq2$ and the fact that 1 is actually a double root of the characteristic polynomial (\charpolyla) points to quadratic degree growth for the iterates of the map. Assuming of course that the express method can actually be extended to such higher order mappings, which is still conjectural.
 
In [\lynett] some of the present authors, together with T. Tamizhmani, studied the Lyness mapping in detail from the point of view of integrability. For example, its growth properties were studied using the approach introduced  by Halburd under the name of Diophantine  integrability [\refdef\dioph].  In this method one considers the iterates  of the mapping for rational initial conditions and with rational values of the parameter and one studies the growth of their arithmetic height $H_n=\max(|p_n|,|q_n|)$ (where $x_n=p_n/q_n$, with $\gcd(p_n,q_n)=1$ .  The Diophantine  integrability criterion introduced by Halburd requires that $\log H_n$ grow no faster than  a polynomial in $n$. In all cases studied in [\lynett] (up to $N=20$) it was found that the limit $(\log H_n)/n^2$ for $n\to \infty$ converges to a finite number, again indicating quadratic growth for the map and hence integrability.
 
Still, the fact that two, admittedly, stringent integrability criteria are satisfied does not constitute a proof of the integrability of the Lyness mapping. It turns out however that we can offer such proof with the help of the bilinear formalism.
In this formalism the first step is to find an adequate ansatz linking the nonlinear variable $x_n$ to a set of  $\tau$-functions. As explained in [\lynett], the derived Lyness map (\zpen) has a confining singularity that is similar to that for (\ztes). The corresponding (confined) singularity pattern easily yields a useful ansatz and it turns out that a single $\tau$-function $\tau_n$ suffices to cast the general mapping in a bilinear form. In fact, it suffices to encode the confined singularity pattern (for general $N\geq2$) in two different ways, one describing the sequence of appearance of zeros in the pattern, the other one describing the appearances of -1 and both yielding the same positions for infinities, in terms of zero values for $\tau_n$:  
$$x_n=A_n {\tau_{n-N}\tau_{n+N+1}\over\tau_{n}\tau_{n+1}}=-1-{\tau_{n-N+1}\tau_{n+N}\over\tau_{n}\tau_{n+1}}.\eqdef\zoct$$
As explained in great detail in [\lynett], it can be seen that for this ansatz to satisfy the discrete derivative of the Lyness mapping it is necessary that $A_n$ be a $(N-1)$-periodic function. 
 The bilinear equation for $\tau_n$ is then obtained by simply equating the two rational expressions in (\zoct), leading to
$$A_n\tau_{n-N}\tau_{n+N+1}+{\tau_{n-N+1}\tau_{n+N}}+\tau_{n}\tau_{n+1}=0, \eqdef\zenn$$ 
which is just a reduction of the Hirota-Miwa equation [\lynett], which is the archetypal integrable system in three discrete dimensions (discrete analogue of the KP equation)  [\refdef\hirota], [\refdef\miwa].
The relation of Lyness to the Hirota-Miwa explains also the quadratic growth of the iterates.  As shown in [\refdef\mase] by T. Mase, the degree of any mapping that can be obtained as a direct reduction of the A-or B-type discrete KP equations, can only grow as $n^\ell$ where $\ell$ is $0, 1$ or $2$. Hence, the degree growth of any non-linear mapping obtained from such a bilinear form through a rational transformations such as those in (\zoct) can only have polynomial growth as well and the Lyness mapping is integrable for all $N$.
\bigskip
3. {\scap Deautonomising the Lyness mapping}
\medskip
As a warm-up (and for reasons that will become clear shortly) we start with a detailed study of the deautonomisation of the Lyness mapping (\ztes) for the case $N=2$. More precisely, we assume that the parameter $a$ entering (\ztes) is now no longer a constant but depends on $n$,
$$x_{n+1}x_{n-1}=a_n+ x_{n}, \eqdef\zdek$$
and require that the singularity pattern of the non-autonomous mapping (\zdek) be identical to that for the autonomous $a_n=a$ case. Again we start with a generic $x_0$ and set $x_1=-a_1$. Iterating, as before, we reproduce the pattern obtained in the autonomous case, $\{-a_1, 0,-{a_2\over a_1},\infty,\infty,-{a_1\over a_2},0, -{a_2 a_7\over a_1}\}$, up to $x_8$ but find that for $x_9$ to depend on $x_0$ a condition on $a_n$ is necessary:
$$a_{n+7}a_n=a_{n+1}a_{n+6}.\eqdef\dena$$
The solution of this condition is $a_n=\kappa\lambda^n$ times a periodic function of period 6. {(It is interesting to remark that Cima and Zafar, who in  [\refdef\cima] studied purely periodic deautonomisations of mapping (\zdek) in which $a_n$ does not contain any non-periodic (secular) contributions, have rigorously shown that the only periodicities in $a_n$ that lead to an integrable mapping are 1, 2, 3 and 6, in perfect agreement with our result)}.
At this point it is useful to introduce two handy periodic functions (that we have been regularly using in publications of ours). The first, $\phi_m(n)$, of period $m$, can be expressed in terms of the roots of unity as
$$ \phi_m(n)=\sum_{l=1}^{m-1} \delta_l^{(m)} \exp\left({2i\pi ln\over m}\right).\eqdef\ddyo$$
(Note that the constant term is absent, the sum starting at $i=1$, involving $m-1$ parameters). Since in this paper we deal with multiplicative equations, we also introduce the `multiplicative' analogue of $\phi_m(n)$, by defining $\varphi_m(n)=\exp(\phi_m(n))$. Note that while $\phi_2(n)+\phi_2(n+1)=0$ we have now $\varphi_2(n)\varphi_2(n+1)=1$ and analogous relations for the higher periods. Similarly we introduce the periodic function $\chi_{2m}(n)$ by
$$ \chi_{2m}(n)=\sum_{\ell=1}^{m} \eta_{\ell}^{(m)} \exp\left({i\pi(2\ell-1)n\over m}\right).\eqdef\dtri$$
which obeys the relation $\chi_{2m}(n+m)+\chi_{2m}(n)=0$. (Note that $\chi_{2m}(n)$ involves just $m$ parameters). Again due to the present multiplicative setting we must introduce $\psi_{2m}=\exp(\chi_{2m}(n))$, an $m$- periodic function which satisfies the relation $\psi_{2m}(n+m)\psi_{2m}(n)=1$. The identity $\varphi_{2N}(n)=\varphi_N(n)\psi_{2N}(n)$, will be useful in what follows.

The solution of (\dena) can now be written as $a_n=\kappa\lambda^n\varphi_6(n)$. However a ternary gauge freedom $x_n\to\psi_6x_n$ is present in (\zdek) and when this spurious freedom is removed the solution becomes finally  $a_n=\kappa\lambda^n\varphi_3(n)\varphi_2(n)$. Equation (\zdek) with this precise expression for $a_n$ is a well-known $q$ discrete Painlev\'e equation.

Before proceeding to the study of higher $N$s, it is interesting to investigate the possibility of late confinements for (\zdek). It turns out that this is indeed possible. The first late confinement corresponds to a singularity pattern $\{-a_1, 0,-{a_2\over a_1},\infty,\infty,-{a_1\over a_2},0,-{a_2 a_7\over a_1},\infty,\infty,-{a_1\over a_2 a_7},0,-{a_2 a_7 a_{12}\over a_1}\}$. The confinement condition needed for $x_{14}$ to depend on $x_0$ instead of becoming infinite, is
$$a_{n+12}a_n=a_{n+11}a_{n+6}a_{n+1}.\eqdef\dtes$$
The corresponding characteristic polynomial,
$$k^{12}-k^{11}-k^6-k+1=(k^2-k+1) (k^{10}-k^8-k^7+k^5-k^3-k^2+1),\eqdef\dpen$$
is the product of an inconsequential cyclotomic factor and a degree 10 polynomial. Following the full-deautonomisation approach we claim that the largest zero of the latter, approximately 1.29348, should give the dynamical degree of the system. The Diophantine method leads to an approximate value in perfect agreement with the one obtained from this characteristic polynomial. 

This is not the only possibility for late confinement for (\zdek). In fact, infinitely many do exist. The next confined singularity pattern can be represented as $\{-a_1, 0,\bol,\infty,\infty,\bol,0,\bol,\infty,\infty,\bol,0,\bol,\infty,\infty,\bol,0, \bol\}$ with confinement condition 
$$a_{n+17}a_n=a_{n+16}a_{n+11}a_{n+6}a_{n+1},\eqdef\dhex$$
leading to the characteristic polynomial
$$k^{17} -k^{16}-k^{11}-k^6-k+1=(k^3+1)(k^{14}-k^{13}-k^{11}+k^{10}-k^7+k^4-k^3-k+1).\eqdef\dhep$$
The largest zero is now approximately 1.31820 and this estimate of the dynamical degree is again verified by the Diophantine approach.

Once the structure of the characteristic polynomial becomes clear it is easy to guess the equation satisfied by the dynamical degree for an infinitely postponed confinement. (Details on how to deal with a infinitely-late confinement can be found in [\refdef\noncon]). To make a long story short we find that the dynamical degree for an infinitely late confinement of the singularity of (\zdek) is the largest root of the equation
$$k^3-k-1=0,\eqdef\doct$$
which we found to be equal to  $\left({1\over2}+\sqrt{{23\over108}}\right)^{1/3}+\left({1\over2}-\sqrt{{23\over108}}\right)^{1/3}$, approximately equal to 1.32472, a value again confirmed by the Diophantine approach for (\zdek) with arbitrary $a_n$. (This value of the dynamical degree is corroborated by the rigorous results in [\cima] where dynamical degrees were calculated for iterates of (\zdek) with the wrong periodicities in $a_n$, i.e. not compatible with integrability).

We now turn our attention to the case of higher $N$. We studied the cases of several low values for $N$, starting from $N=3$ and all of them exhibited the same behaviour. Starting from generic $x_0, x_1,\cdots, x_{N-2}$ and choosing $x_{N-1}=x_*$ such that $x_N=0$, we found that after the two expected infinities for $x_{2N}$ and $x_{2N+1}$, an extra infinity appeared in $x_{2N+2}$ instead of a non-zero finite value as for the autonomous case. In fact, it turns out that the only way to make this infinity disappear is to require that $a_n$ be constant. Hence, it does not appear to be possible to deautonomise the Lyness mapping for any $N$ higher than 2. Still, it is interesting to study the non-confining singularity pattern produced by this deautonomisation attempt. For $N=3$ we found the pattern
$\{x_*, 0,\bol,\bol,\infty,\infty,\infty,\bol,\bol,0,\bol,\bol,\infty,\infty,\infty\bol,\cdots\}$ in which the basic pattern $\{0,\bol,\bol,\infty,\infty,\infty,\bol,\bol\}$ repeats indefinitely. For $N=4$ the basic repeating pattern is $\{0,\bol,\bol,\bol,\infty,\infty,\infty,\infty,\bol,\bol,\bol\}$, while for $N=5$ we find $\{0,\bol,\bol,\bol,\bol,\infty,\infty,\infty,\infty,\infty,\bol,\bol,\bol,\bol\}$. We therefore surmise that for general $N$ the non-confined singularity pattern will be given by the infinite repetition of the basic pattern
$$\{0,(N-1\ {\rm finite\ values}\ \bol),(N\ {\rm values}\ \infty),(N-1\ {\rm finite\ values}\ \bol)\}.$$
Based on the structure of this repeating pattern it is straightforward to obtain a characteristic equation that would give the value of the dynamical degree of a (non-integrable) non-autonomous mapping for $N>2$. We shall not go into the details of the derivation (which follows the method of Halburd [\refdef\rod], as we explained in [\express]) and just give the result. We claim that the dynamical degree is given by the largest root of the equation
$$k^{2N}-k^{2N-1}-k^N+1=0.\eqdef\denn$$
For $N=3,4,5$ we find the (approximate) values 1.3247, 1.2964, and 1.2686, respectively. (Note that for $N=3$ the value of this root is exactly that obtained for the infinitely-late confinement for  $N=2$. In fact, the characteristic equation (\denn) contains (\doct) as a factor both for $N=2$ and $N=3$). Using the Diophantine approach we computed the dynamical degrees of a non-autonomous Lyness mapping (for an arbitrary function for $a_n$) for $N=3,4$ and 5, and we found that the dynamical degree obtained by the direct calculation, after 50 iterations, coincided with the largest root of the characteristic equation with a $10^{-4}$ precision. From the values obtained for $N=3,4,5$ it is clear that the largest root of (\denn) diminishes with increasing $N$. A rough estimate of the said root gives a behaviour $1+\log(N)/N$, with a better approximation being $1+(\log(N-1)-\log(\log(N-1)))/(N-1)$.
\bigskip
4. {\scap Deautonomising the derivative Lyness mapping}
\medskip
While the study of the higher order non-autonomous Lyness mappings for $N\geq3$ led to some interesting results, it remains that the fact that these mappings cannot be deautonomised while preserving their integrability is somewhat frustrating. We therefore turn to a different approach and investigate the possibility of deautonomising their derivative form (\zpen) instead. We thus introduce the mapping
$$ x_{n+N}(p_n+x_n)=x_{n+1}(q_{n+N+1} +x_{n+N+1}),\eqdef\ddek$$
where $p_n$ and $q_n$ are functions the form of which will be determined by the application of the confinement criterion, i.e. by requiring that the nonautonomous mapping still leads to the confined singularity pattern $\{0,(N-1 {\rm\ finite\ values\ \bol}),\infty,\infty,(N-1 {\rm\ finite\ values\ \bol}),0, (N-1 {\rm\ finite\ values}\ *)\}$ we find in the autonomous case (here $\bol$ is used to indicate finite and non-zero expressions in $p_n, q_n$ and the free initial conditions, and $*$ to indicate a succession of such functions that are not functionally independent as functions of the (free) initial conditions). 

We start by examining the 4th order mapping we obtain in the $N=3$ case. Starting from generic $x_0,x_1$ and $x_2$, and $x_3=0$, we find that subsequent iterates match the required singularity pattern, but that $x_8$ is also infinite unless $q_n=p_n$. We apply this first confinement constraint and pursue the iterations. We find that $x_9$ and $x_{10}$ are finite but that $x_{10}$ is not 0 unless $p_n$ satisfies the condition
$$p_{n+7}p_n=p_{n+6}p_{n+1}.\eqdef\vena$$
Implementing this condition as well we find that the resulting mapping does indeed possess the desired confined singularity pattern.
Integrating (\vena) is straightforward and leads to $p_n=\kappa\lambda^n\varphi_6(n)$. However, since equation (\ddek) is invariant under a gauge transformation $x_n\to\psi_{2N}x_n$ we find, after removing a spurious factor $\psi_6(n)$, that the coefficients in (\ddek) must take the form $q_n=p_n=\kappa\lambda^n\varphi_3(n)$.

What happens if we decide to take $q_n\ne p_n$? In that case the singularity pattern becomes unconfined $\{0,\bol,\bol,\infty,\infty,\infty,\bol,\bol,\bol,\infty,\infty,\infty,\cdots\}$, where the pattern $\{\infty,\infty,\infty,\bol,\bol,\bol\}$ repeats indefinitely. The resulting characteristic equation in this case is $k^6-k^3-k^2-k-1=0$ with largest root, approximately,  1.3803. If we decide to take  $q_n=p_n$ but ignore the second confinement condition, equation (\vena), we find a singularity pattern $\{0,\bol,\bol,\infty,\infty,\bol,\bol,\bol,\infty,\infty,\cdots\}$ with the pattern $\{\infty,\infty,\bol,\bol,\bol\}$ repeating indefinitely. The characteristic equation is now $k^5-k^2-k-1=0$, which contains (\doct) as a factor  (the value of the dynamical degree of being confirmed by a Diophantine calculation).

The case $N=2$ has also a similar deautonomisation. Starting with generic $x_0,x_1$ and $x_2=0$, subsequent iterates match the desired pattern but $x_6$ is infinite unless $q_n=p_n$. Iterating further under this constraint we find that $x_7$ is finite but non-zero unless $p_n$ satisfies the condition
$$p_{n+5}p_n=p_{n+4}p_{n+1}.\eqdef\vdyo$$
Under this condition the mapping then automatically possess the required confined singularity pattern.
The solution of (\vdyo) is $p_n=\kappa\lambda^n$ with, moreover, a period-4 freedom, but after the removal of the freedom due to gauge we have finally $q_n=p_n=\kappa\lambda^n\varphi_2(n)$. 

We have studied several cases corresponding to higher values of $N$. For all of them a first confinement constraint is $q_n=p_n$. Imposing this constraint we obtain, after a certain number of iterations, that $p_n$ must satisfy a second constraint which, we surmise, has the form
$$p_{n+2N+1}p_n=p_{n+2N}p_{n+1}.\eqdef\vtri$$
Once gauge freedom has been removed, one finds that $q_n=p_n=\kappa\lambda^n\varphi_N(n)$. 

As in the case $N=3$ above we can decide to take $q_n\ne p_n$ or, when $q_n=p_n$, to ignore the second confinement condition. The resulting singularity patterns are obtained by the infinite repetitions of a simple basic pattern, which in the former case consists of $N$ infinities followed by $N$ finite values, while in the latter case it starts with 2 successive infinities, followed by $N$ finite values, followed by $N-1$ infinities and $N$ finite values. It is straightforward to obtain the characteristic equation in both cases. In the former we find
$$k^{N+1}-k^N-1=0,\eqdef\vtes$$
and in the latter $k^{3N+2}-k^{3N+1}-k^{2N+2}+k^{2N}-k^N+1=0$, which by dividing by factor $(k^{N+1}-1)$ can be brought to the form
$$k^{2N+1}-k^{2N}-k^{N+1}+k^N-1=0.\eqdef\vpen$$
For (\vtes) a rough estimate of the largest root is  $1+\log(N)/N$, while for (\vpen) the same reasoning gives $1+\log(2N)/(2N)$.

Taking $q_n=p_n$ in (\ddek) allows one to integrate once (and this, in fact, independently of the explicit expression for $p_n$). We start from (\ddek) rewritten as
$${p_{n+N+1} +x_{n+N+1}\over p_n+x_n}={x_{n+N}\over x_{n+1}}.\eqdef\vhex$$
We remark that in the left-hand side the indices of the numerator and denominator are at a distance $N+1$ while in the right-hand side the distance is $N-1$. 
Next we multiply the numerator and denominator of the left-hand side by $(p_{n+N} +x_{n+N})(p_{n+N-1} +x_{n+N-1})\cdots(p_{n+1} +x_{n+1})$ while for the right-hand side the corresponding multiplier is $x_{n+N-1}x_{n+N-2}\cdots x_{n+2}$. It is now straightforward to integrate the resulting equation. We find
$$(p_{n+N} +x_{n+N})(p_{n+N-1} +x_{n+N-1})\cdots(p_n+x_n)=Cx_{n+N-1}x_{n+N-2}\cdots x_{n+1}\eqdef\vhep$$
It is interesting at this point to give the explicit expressions of these mappings  for a few low-N cases. In order to write the equations in a more familiar form we introduce the variable $X_n=x_n+p_n$. 
For $N=2$ we have 
$$X_{n+1}X_{n-1}=C{X_n-p_n\over X_n}\eqdef\voct$$
which, with $p_n=\kappa\lambda^n\varphi_2(n)$, is a well-known [\refdef\martin] discrete Painlev\'e equation. 
For $N=3$ we start by remarking that the distance of indices in both sides of (\vhex) is even. (And this remark is valid for all odd $N$s). Thus it suffices to multiply by only one out of the two factors we used above in order to be able to integrate. The integration now introduces a period-2 function in lieu of the integration constant we had before, but given the parity of the factors this is precisely the freedom that can be removed by the appropriate choice of the gauge. Finally, we find
$$X_{n+1}X_{n-1}=C(X_n-p_n),\eqdef\venn$$
with $p_n=\kappa \lambda^n\varphi_6(n)$, which is nothing but the discrete Painlev\'e equation that we obtained in Sec. 3 by deautonomising the $N=2$ case of the Lyness mapping in its original form.

For $N=4$ we find
$$X_{n+2}X_{n+1}X_nX_{n+1}X_{n+2}=C(X_{n+1}-p_{n+1})(X_n-p_n)(X_{n-1}-p_{n-1}),\eqdef\vdek$$
while for $N=5$ we again find a 4th-order mapping:
$$X_{n+2}X_nX_{n+2}=C(X_{n+1}-p_{n+1})(X_{n-1}-p_{n-1}).\eqdef\tena$$
We claim that (with the appropriate $p_n$) equation (\vhep) for $N$ even, and integrated once more for $N$ odd,  is a possible integrable deautonomisation of the higher $N$ Lyness mapping, a deautonomisation that was impossible to obtain by the direct deautonomisation of (\ztes).
\bigskip
5. {\scap The mysterious affair of the derivative Lyness mapping for  $N=2$}
\medskip
At this point the careful reader has certainly noticed that we started our presentation in the previous section with the $N=3$ case and, once the results were obtained for that case, we proceeded to deal with the $N=2$ case.  The reason for this will become clear in this section. We start from 
$$x_{n+2}(p_n+x_n)=x_{n+1}(q_{n+3} +x_{n+3}),\eqdef\tdyo$$
and study its singularity pattern starting with generic $x_0,x_1$ and $x_2=0$. We obtain the sequence of values finite, infinite, infinite, finite, as expected but then $x_7$ is finite, instead of 0, thus leading to an infinite value for $x_8$. In order to remedy this a constraint must be imposed on $p_n$ and $q_n$:
$$p_{n+1}p_n-q_{n+5}p_n-q_{n+1}p_n+q_{n+1}q_{n+4}=0.\eqdaf\ttri$$
However, iterating further we find that $x_9$ becomes infinite unless a second constraint is introduced
$$p_{n+4}q_{n+1}-p_{n+1}p_n+q_{n+1}p_n-q_{n+1}q_{n+4}=0.\eqno(\ttri \rm b)$$
Once the two constraints are satisfied the singularity becomes indeed confined with the expected pattern $\{ 0,\bol,\infty,\infty,\bol,0, *, *\}$. 

Solving the confinement constraints is particularly interesting. Taking the sum of (\ttri a) and (\ttri b) we find that  $p_n$ and $q_n$ obey the simpler relation $p_{n+4}q_{n+1}=q_{n+5}p_n$ which means that 
$p_n=z_nq_{n+1}$ with $z_{n+4}=z_n$. Next we introduce  $w_n=p_{n+1}-q_{n+1}$ and from (\ttri b), rewritten as $p_n w_n = q_{n+1} w_{n+3}$, we find that $w_{n+3}=z_n w_n$. Eliminating $z_n$, we obtain for $w_n$ the equation
$$w_{n+7}w_n=w_{n+3}w_{n+4}.\eqdef\ttes$$
The solution of (\ttes) is $w_n=\kappa\lambda^n$ multiplied by two periodic terms with periods 3 and 4.

Rewriting the relation $w_n=p_{n+1}-q_{n+1}$ in terms of $w_n$ using $z_n=w_{n+3}/w_n$ we have $w_{n+4}q_{n+2}-w_{n+1}q_{n+1}=w_nw_{n+1}$. Introducing  $Q_n=w_{n+1}w_{n+2}w_{n+3}q_{n+1}$ we find that $Q_n$ must obey the equation
$$Q_{n+1}-Q_n=w_nw_{n+1}w_{n+2}w_{n+3},\eqdef\tpen$$
that can be integrated by a simple quadrature.

We find $q_{n+1}=\varphi_4(n)(\sigma\varphi_3(n)\lambda^n+\mu\lambda^{-3n})$ and  $p_n=\lambda^3\varphi_4(n-1)(\sigma\varphi_3(n)\lambda^n+\mu\lambda^{-3n})$, where $\mu$ is a free constant and $\sigma$ is proportional to $\kappa$. This is an original non-autonomous form
with two different exponential terms in the same parameter, which is something that we have never encountered before.  (Whether this is a general feature of third- and higher-order equations remains to be seen).

A consequence of the presence of two exponentials in $q_n$ and $p_n$ is that we can turn the $n$ dependence in these parameters into an additive one, by taking the limit $\lambda\to1$, without changing the functional form of the equation. First, we remark that $\sigma\varphi_3(n)$ becomes $\sigma+\phi_3(n)$ by taking the appropriate limits on the parameters of the periodic function. Next we introduce $\lambda=1+\delta$ and take $\delta\to0$. We assume that $\sigma$ and $\mu$ diverge as $1/\delta$, obeying the relations $\sigma-3\mu=\alpha/\delta$ and $\sigma+\mu=\beta$. Taking the limit $\delta\to 0$ we find for $q_n$ and $p_n$ the expressions $q_{n+1}=\varphi_4(n)(\alpha n+\beta+\phi_3(n))$ and $p_n=\varphi_4(n-1)(\alpha n+\beta+\phi_3(n))$.

But what is even more interesting is the case of late confinement of (\tdyo).
The expected pattern is now $\{ 0,\bol,\infty,\infty,\bol,\bol,\infty,\infty,\bol,0\},$ from which we can infer the characteristic equation $k^9-k^7-k^6-k^3-k^2+1=0$ with largest root approximately 1.42501.
Obtaining the confinement constraints requires some extensive calculations. Performing them we find the two confinement conditions
$$\displaylines{p_{n + 8}p_{n + 1}p_{n}q_{n+1} - p_{n + 8}p_{n}q_{n + 5}q_{n+1} - p_{n + 8}p_{n}q_{n+1}^2 + p_{n + 8}q_{n + 4}q_{n+1}^2\hfill\cr\hfill
 + p_{n + 4}p_{n}q_{n +9}q_{n+1} - p_{n + 1}p_{n}^2q_{n + 9} + p_{n}^2q_{n + 9}q_{n+1} - p_{n}q_{n + 9}q_{n + 4}q_{n+1}=0,\quad\eqdisp\thex\cr}$$
 and
$$\displaylines{p_{n + 8}p_{n + 1}p_{n}q_{n+1} - p_{n + 8}p_{n}q_{n + 5}q_{n+1} - p_{n + 8}p_{n}q_{n+1}^2 + p_{n + 8}q_{n + 4}q_{n+1}^2 + p_{n + 5}p_{n + 4}p_{n}q_{n+1}- p_{n + 5}p_{n + 1}p_{n}^2 \hfill\cr\hfill + p_{n + 5}p_{n}^2q_{n+1} - p_{n + 5}p_{n}q_{n + 4}q_{n+1}+ p_{n + 4}p_{n + 1}p_{n}q_{n+1} - p_{n + 4}p_{n}q_{n + 5}q_{n+1} - p_{n + 4}p_{n}q_{n+1}^2 + p_{n + 4}q_{n + 4}q_{n+1}^2\hfill\cr\hfill- p_{n + 1}^2p_{n}^2 + p_{n + 1}p_{n}^2q_{n + 5}  + 2p_{n + 1}p_{n}^2q_{n+1} - p_{n + 1}p_{n}q_{n + 8}q_{n+1}- 2p_{n + 1}p_{n}q_{n + 4}q_{n+1} - p_{n}^2q_{n + 5}q_{n+1} - p_{n}^2q_{n+1}^2   \hfill\cr\hfill+ p_{n}q_{n + 5}q_{n + 8}q_{n+1}+ p_{n}q_{n + 8}q_{n+1}^2 + p_{n}q_{n + 5}q_{n + 4}q_{n+1}+ 2p_{n}q_{n + 4}q_{n+1}^2 - q_{n + 8}q_{n + 4}q_{n+1}^2 - q_{n + 4}^2 q_{n+1}^2=0.\quad\eqdisp\thep\cr}$$
The full-deautonomisation method stipulates that the confinement conditions lead directly to the value of the dynamical degree. In all  cases studied up to now the deautonomisation conditions were simple linear or multiplicative equations and the dynamical degree could be obtained through the study of their characteristic equation. Here we are in presence of a system that is non-linear and, in fact, non-integrable. Still, it can lead to the value of dynamical degree. 

To this end we study the growth of the solutions of the system, introducing initial conditions where $p_i, q_i$, $i=0,6$ are generic, and, for simplicity,  taking $p_7$ generic and $q_7=\mu$. We compute the degrees in $\mu$ of the iterates of (\thex) and (\thep) and, starting at $n=8$, find the sequence 1, 1, 1, 2, 3, 4, 6, 10, 13, 18, 27, 38, 54, 75, 111, 158, 225, 321, 457, 652, $\cdots$. The ratio of 457/321 is 1.42368, the ratio 652/457 is 1.42670 and  taking their geometric mean we obtain 1.42519, very close to the dynamical degree 1.42501 inferred above. As a further verification, we studied the degree growth of (\thex), (\thep) through the Diophantine approach and obtained a dynamical degree approximately equal to 1.4256, confirming the result of the characteristic equation and that of the direct calculation. Thus the dynamical degree is indeed given by the growth of the variables $p_n$ and $q_n$ when they obey the confinement conditions.
\bigskip
6. {\scap Conclusion}
\medskip
The Lyness mapping is a simple integrable system, the deautonomisation of which however turned out to be particularly interesting. Its integrable character was already established in [\lynett] where we have shown that it can be obtained as a reduction of the Hirota-Miwa integrable lattice equation. The present work was devoted to the  study of the deautonomisation of this mapping in two different forms. As we have already shown, the deautonomisation process allows to accentuate the deep relationship between the singularity structure of the solutions of an equation and its integrable character, the latter being deduced from the value of the dynamical degree.  Deautonomising the Lyness mapping under its customary form turned out to be possible only in the case $N=2$. However, by considering the derivative form of the mapping, we were able to produce an (admittedly tailored to this special form) deautonomisation for every value of $N$. In this way we presented a method to construct integrable equations of arbitrary (even) order which include well-known $q$-discrete Painlev\'e equations at $N=2$ and 3.

Studying cases of `late' confinement for both the $N=2$ and higher $N$ cases (the latter through the derivative form), the deautonomisation allowed once more to confirm the predictions of the `full-deautonomisation' approach, successfully comparing the dynamical degree thus obtained with the value given by a direct Diophantine calculation. The deautonomisation of the $N=2$ case, however, reserved some surprises. First, instead of the situation familiar in the case of $q$-Painlev\'e equations where the $n$-dependence in the parameters enters through a single exponential, here we were in the presence of two exponential terms in the same parameter. And as a consequence of this we found that, remarkably, a deautonomisation where $n$ enters linearly was also possible for the same functional form of the equation. But the most interesting result was the one obtained by considering the late confinement in this case. We obtained a system of two nonlinear, non-integrable equations, where the dynamical degree did not appear explicitly (as in all previously studied cases) as the largest root of some characteristic equation. Here we had to study the growth of the solutions of the nonlinear system, with respect to some initial condition. It turned out that the dynamical degree obtained from the growth of the solutions of the confinement conditions coincides with the dynamical degree of the solution of the mapping (obtained using Halburd's method and confirmed by the Diophantine approach).

This result is one more validation of the full-deautonomisation approach as a reliable integrability criterion. At this point we cannot resist the temptation to quote the `prophetic' statement of Gambier [\refdef\gambier], who, while studying differential equations from the point of view of the Painlev\'e property, remarked:

{\sl Je rencontrais des syst\`emes de conditions diff\'erentielles dont
l'int\'egration \'etait, quoiqu'au fond bien simple, assez difficile \`a apercevoir.
Par un m\'ecanisme qui est g\'en\'eral, mais qui \'etait difficile \`a pr\'evoir, la
r\'esolution de ce premier probl\`eme, int\'egration des conditions, est intimement
li\'ee \`a l'int\'egration de l'\'equation diff\'erentielle elle-m\^eme}.

(I found myself confronted with systems of differential conditions whose integration, though in itself quite simple, was nevertheless difficult to detect. By means of a general mechanism, yet one which it would have been hard to foresee, the solution of this first problem, the integration of the conditions, is intimately related to the integration of the differential equation itself).

This is exactly what full-deautonomisation is about, and the study of the Lyness mapping offers a novel and interesting setting for its application.

\bigskip
{\scap Acknowledgements}
\medskip
RW would like to thank the Japan Society for the Promotion of Science (JSPS)
for financial support through the KAKENHI grant 23K22401.

\bigskip
{\scap References}
\medskip
\item{[\kruskal]} N.J. Zabusky and M.D. Kruskal, {\sl Interaction of "Solitons" in a Collisionless Plasma and the Recurrence of Initial States}, Phys. Rev. Lett. {\bf 15} (1965) 240.
\item{[\qrt]} G.R.W. Quispel, J.A.G. Roberts and C.J. Thompson, {\sl Integrable mappings and soliton equations II}, Physica D {\bf 34} (1989) 183.
\item{[\veselov]} A.P. Veselov, {\sl Growth and integrability in the dynamics of mappings}, Commun. Math. Phys. {\bf 145} (1992) 181.
\item{[\baxter]} R.J. Baxter, {\sl Exactly Solved Models in Statistical Mechanics}, Associated Press, London (1982), p. 471.
\item{[\autoqrt]} A. Ramani, S. Carstea, B. Grammaticos and Y. Ohta, {\sl On the autonomous limit of discrete Painlev\'e equations}, Physica A  {\bf 305} (2002) 437.
\item{[\iatrou]} A. Iatrou and J.A.G. Roberts, {\sl Integrable mappings of the plane preserving biquadratic invariant curves II}, Nonlinearity {\bf 15} (2002) 459. 
\item{[\sincon]} B. Grammaticos, A. Ramani and V. Papageorgiou, {\sl Do Integrable Mappings Have the Painleve Property?}, Phys. Rev. Lett. {\bf 67} (1991) 1825.
\item{[\tremblay]} A. Ramani, B. Grammaticos and S. Tremblay, {\sl Integrable systems without the Painlev\'e property}, J. Phys. A: Math. Gen. {\bf 33} (2000) 3045.
\item{[\hiv]} J. Hietarinta and C. Viallet, {\sl Singularity confinement and chaos in discrete systems}, Phys. Rev. Lett. {\bf 81} (1998) 325.
\item{[\diller]} J. Diller and C. Favre, {\sl Dynamics of bimeromorphic maps of surfaces}, Amer. J. Math. {\bf 123} (2001) 1135.
\item{[\redemp]} A. Ramani, B. Grammaticos, R. Willox, T. Mase and M. Kanki, {\sl The redemption of singularity confinement}, J. Phys. A: Math. Theor. {\bf 48} (2015) 11FT02.
\item{[\royal]} T. Mase, R. Willox, B. Grammaticos and A. Ramani, {\sl Deautonomisation by singularity confinement: an algebro-geometric justification}, Proc. R. Soc. A {\bf 471} (2015) 20140956.
\item{[\redeem]} B. Grammaticos, A. Ramani, R. Willox, T. Mase and J. Satsuma, {\sl Singularity confinement and full-deautonomisation: a discrete integrability criterion},  Physica D {\bf 313} (2015) 11.
\item{[\stokes]} A. Stokes, T. Mase, R. Willox and B. Grammaticos, {\sl Deautonomisation by singularity confinement and degree growth}, J. Geom. Anal. {\bf 35} (2025) 65.
\item{[\higher]} A. Ramani, B. Grammaticos, A.S. Carstea and R. Willox, {\sl Obtaining the growth of higher order mapping through the study of singularities}, J. Phys. A: Math. Theor. {\bf 58} (2025) 115201.
\item{[\lyness]} R. C. Lyness, {\sl Mathematical Notes 1581. Cycles}, The Mathematical Gazette {\bf 26} (1942) 207. 
\item{[\lynesstoo]} R. C. Lyness, {\sl Mathematical Notes 1847. Cycles}, The Mathematical Gazette {\bf 29} (1945) 231. 
\item{[\kocicetal]} V.L. Kocic, G. Ladas and I.W. Rodrigues, {\sl On recursive sequences}, J. Math. Anal. Appl. {\bf 173} (1993) 127. 
\item{[\tranetal]} D.T. Tran, P.H. van der Kamp and G.R.W. Quispel, {\sl Sufficient number of integrals for the pth-order Lyness equation}, J. Phys. A: Math. Theor. {\bf 43} (2010) 302001.
\item{[\groveladas]} E.A. Grove and G. Ladas, {\sl Periodicity in nonlinear difference equations}, CUBO, Matem\'atica Educacional Vol. {\bf 4} No. 1 (2002) 192. 
\item{[\lynett]} B. Grammaticos, A. Ramani and T. Tamizhmani, {\sl Investigating the integrability of the Lyness mappings}, J. Phys. A: Math. Theor. {\bf 42} (2009) 454009.
\item{[\express]} A. Ramani, B. Grammaticos, R. Willox and T. Mase, {\sl Calculating algebraic entropies: an express method}, J. Phys. A: Math. Theor. {\bf 50} (2017) 185203.
\item{[\dioph]} R.G. Halburd, {\sl Diophantine Integrability}, J. Phys. A: Math. Gen. {\bf 38} (2005) L263.
\item{[\hirota]} R. Hirota, {\sl Discrete Analogue of a Generalized Toda Equation}, J. Phys. Soc. Jpn. {\bf 50} (1981) 3785.
\item{[\miwa]} T. Miwa, {\sl On Hirota's difference equations}, Proc. Japan. Acad. {\bf 58} (1982) 9.
\item{[\mase]} T. Mase, {\sl Investigation into the role of the Laurent property in integrability}, J. Math. Phys. {\bf 57} (2016) 022703.
\item{[\cima]} A. Cima and S. Zafar, {\sl Integrability and algebraic entropy of $k$-periodic non-autonomous Lyness recurrences}, J. Math. Anal. Appl. {\bf 413} (2014) 20.
\item{[\noncon]} A. Ramani, B. Grammaticos, R. Willox, T. Mase and J. Satsuma, {\sl Calculating the algebraic entropy of mappings with unconfined singularities}, J. Integr. Sys. {\bf 3} (2018) 1.
\item{[\rod]}  R.G. Halburd, {\sl Elementary exact calculations of degree growth and entropy for discrete equations}, Proc. R. Soc. A {\bf 473} (2017) 20160831.
\item{[\martin]} M.D. Kruskal, K.M. Tamizhmani, B. Grammaticos and A. Ramani, {\sl Asymmetric discrete Painlev\'e equations} , Reg. Chaot. Dyn. {\bf 5} (2000) 273.
\item{[\gambier]} B. Gambier, {\sl Sur les \'equations diff\'erentielles du second ordre et du premier degr\'e dont l'int\'egrale g\'en\'erale est \`a points critiques fixes}, Acta Math. {\bf 33} (1909) 1.
\end

{\bf Should we give an Appendix for Halburd's method and the infinite-late case?}

\end